\documentclass[]{elsarticle}
\usepackage{booktabs} 

\usepackage{url}
\usepackage{multirow}
\usepackage{boxedminipage}
\usepackage{bm}
\usepackage{graphicx}
\usepackage{color,soul}
\usepackage[font=small,font=bf,skip=5pt]{caption}
\usepackage{subfigure}

\begin{document}


	\title{Can the Optimizer Cost be Used to Predict Query Execution Times?}

	\author[ak]{Anthony Kleerekoper\corref{cor1}}
	\ead{a.kleerekoper@mmu.ac.uk}
	\author[man]{Javier Navaridas}
	\ead{javier.navaridas@manchester.ac.uk}
	\author[man]{Mike Lujan}
	\ead{mikel.lujan@manchester.ac.uk}

	\cortext[cor1]{Corresponding author}
	
	\address[ak]{SCMDT, Manchester Metropolitan University, Manchester, UK}
	\address[man]{School of Computer Science, University of Manchester, UK}

\begin{abstract}
Predicting the execution time of queries is an important problem with applications in scheduling, service level agreements and error detection. During query planning, a cost is associated with the chosen execution plan and used to rank competing plans. It would be convenient to use that cost to predict execution time, but it has been claimed in the literature that this is not possible. In this paper, we thoroughly investigate this claim considering both linear and non-linear models. We find that the accuracy using more complex models with only the optimizer cost is comparable to the reported accuracy in the literature.
The most accurate method in the literature is nearest-neighbour regression which does not produce a model. The published results used a large feature set to identify nearest neighbours. We show that it is possible to achieve the same level of accuracy using only the cost to identify nearest neighbours. Using a smaller feature set brings the advantages of reduced overhead in terms of both storage space for the training data and the time to produce a prediction.
\end{abstract}


	
\maketitle


\section{Introduction}

\begin{figure}[t]
	\centering
	\includegraphics[width=\columnwidth,trim=4 4 4 4,clip]{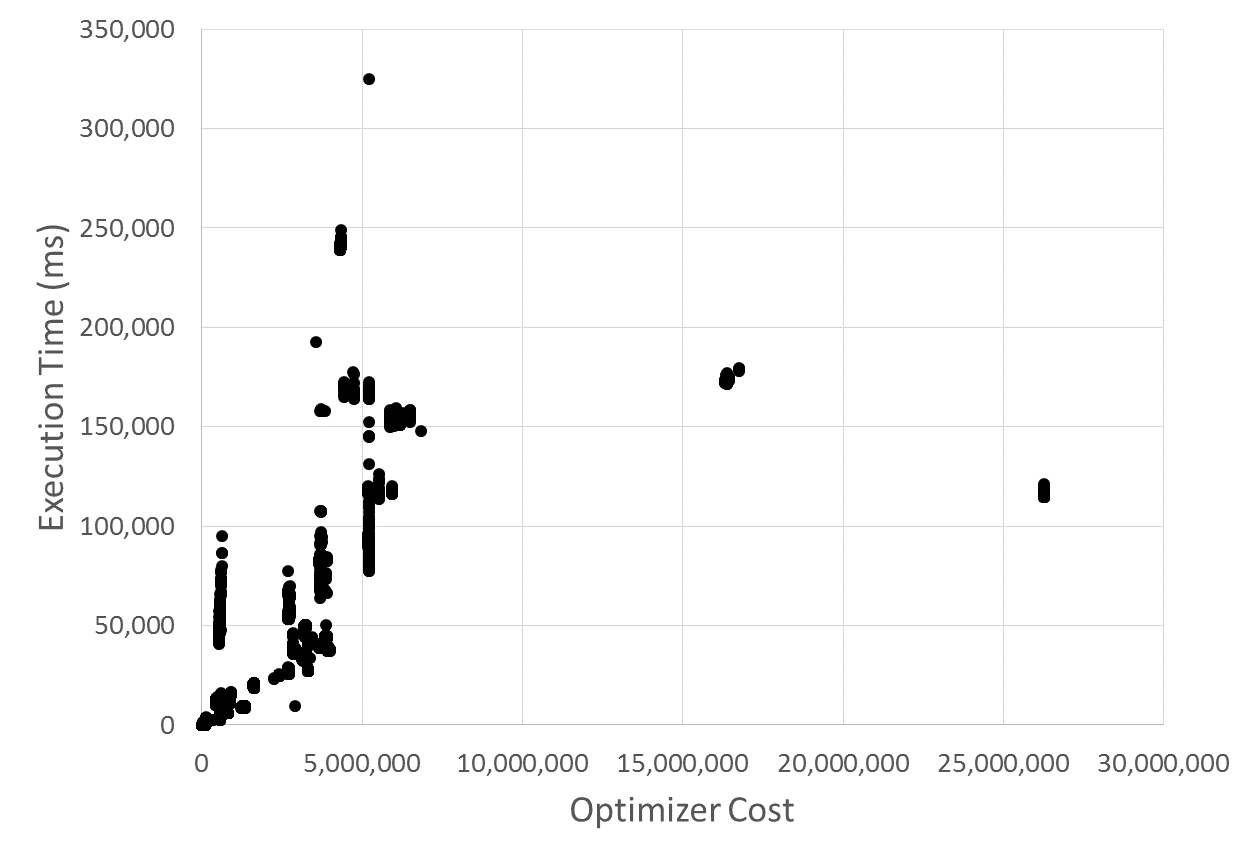}
	\caption{Plotting the execution time of queries against their optimizer cost shows that the cost is not an accurate prediction of execution time}
	\label{fig:CostTimeAll}
\end{figure}

Being able to predict how long queries are going to take to run on a given system is an important problem that has received significant interest in recent years \cite{Gupta2008,Ganapathi2009,Akdere2012,Li2012,Wu2013,Singhal2016}. Known as Query Performance Prediction (QPP), accurately predicting the run-time of SQL queries has three main applications. It can be used to inform scheduling of concurrently executing queries, to guide Service Level Agreements for Database-as-a-Service providers and to identify errors when a query is taking significantly longer to execute than predicted. 

The problem of predicting query execution time is linked to the fundamental problem of query optimization. When a query is executed, the DBMS must create a query execution plan. For any non-trivial query, there will be multiple plans that are equivalent in their output but may differ in their execution time. To decide between alternative plans, the query optimizer calculates and assigns a cost to each plan in order to select the plan with the lowest cost. 

Since the aim of the optimizer is to identify the query plan that will have the shortest execution time, the optimizer's cost itself should be a prediction of the execution time \cite{Selinger1979}. This is an extremely challenging problem and, in practice, the optimizer cost is used only to rank competing plans in terms of their execution time which is sufficient to identify the optimal plan. In fact, sometimes the optimizer cost is not even expressed as a unit of time. For example the PostgreSQL documentation states that ``the cost variables \dots are measured on an arbitrary scale. Only their relative values matter.''\footnote{\url{https://www.postgresql.org/docs/9.6/static/runtime-config-query.html}} 

Figure \ref{fig:CostTimeAll} shows the execution time of queries plotted against the optimizer cost for our three benchmarks (see Section \ref{sec:Benchmarks} for details on the benchmarks). From the graph, it is immediately apparent that the optimizer cost is not an accurate prediction of execution time.

Although the optimizer cost is not itself a prediction of execution time, it would be convenient if some function of the cost were an accurate prediction. This is because the optimizer cost is provided along with the chosen execution plan with no additional overhead. If (reasonably simple) post-processing of the cost was effective, then prediction could be done quickly. However, the consensus in the literature is that no such function exists \cite{Ganapathi2009,Li2012,Akdere2012}. Wu \textit{et al.} summarised the relevant literature, claiming that ``It is clear from this previous work that post-processing the optimizer cost estimate is not effective'' \cite{Wu2013}.

Because the cost was deemed unusable for prediction, alternative methods were proposed. Most alternatives were based on using more features about the query plan than just the cost \cite{Ganapathi2009,Li2012,Akdere2012}. Wu \textit{et al.} proposed a method that could tune the cost parameters in advance so that the cost would become an accurate prediction by itself \cite{Wu2013}. They had to include significant overhead to improve cardinality estimates and even so achieved limited success relative to the post-processing methods.

We note that the consensus against post-processing the optimizer cost is based on limited evidence. Specifically, the claim is based on results in which the post-processing is limited to linear regression. Moreover, of the three studies that considered post-processing the cost, only one presented quantified results \cite{Akdere2012}. By contrast, the alternatives include more complex non-linear models such as SVR and regression trees.

In this paper, we thoroughly investigate the question of post-processing the optimizer cost. Our contributions are:

\begin{itemize}
\item We repeat the experiments from the literature and confirm that linear regression is not effective (Section \ref{sec:EvidenceAgainst}). 
\item We consider a number of alternative post-processing methods and show that some of them produce similar accuracy to methods published in the literature. The methods we consider are:
	\begin{enumerate}
	 \item operator-level linear regression (Section \ref{sec:OpLevelLinReg})
	 \item regression assuming a power-law relationship (Section \ref{sec:PowerLaw}) 
	 \item non-linear regression with SVR (Section \ref{sec:SVR})
	\end{enumerate}
\item We show that when using the non-parametric nearest-neighbour regression method, the optimizer cost is as effective as a larger feature set (Section \ref{sec:KNN}). 
\end{itemize}

We note that nearest-neighbour regression is the most accurate method in the literature. We also note that, unlike the other methods, it can also be used to predict other resources used by a query because the method is essentially a large lookup table. Being able to predict with just the optimizer cost means we can reduce the amount of memory required for this method and also reduce the lookup times. In fact, the paper originally proposing nearest-neighbour regression used such a large feature set that they felt the need to include dimensionality reduction as a pre-processing step. Using only the optimizer cost makes this unnecessary.

\section{Background}

\begin{figure}[thp]
	\centering
	\includegraphics[width=0.9\columnwidth]{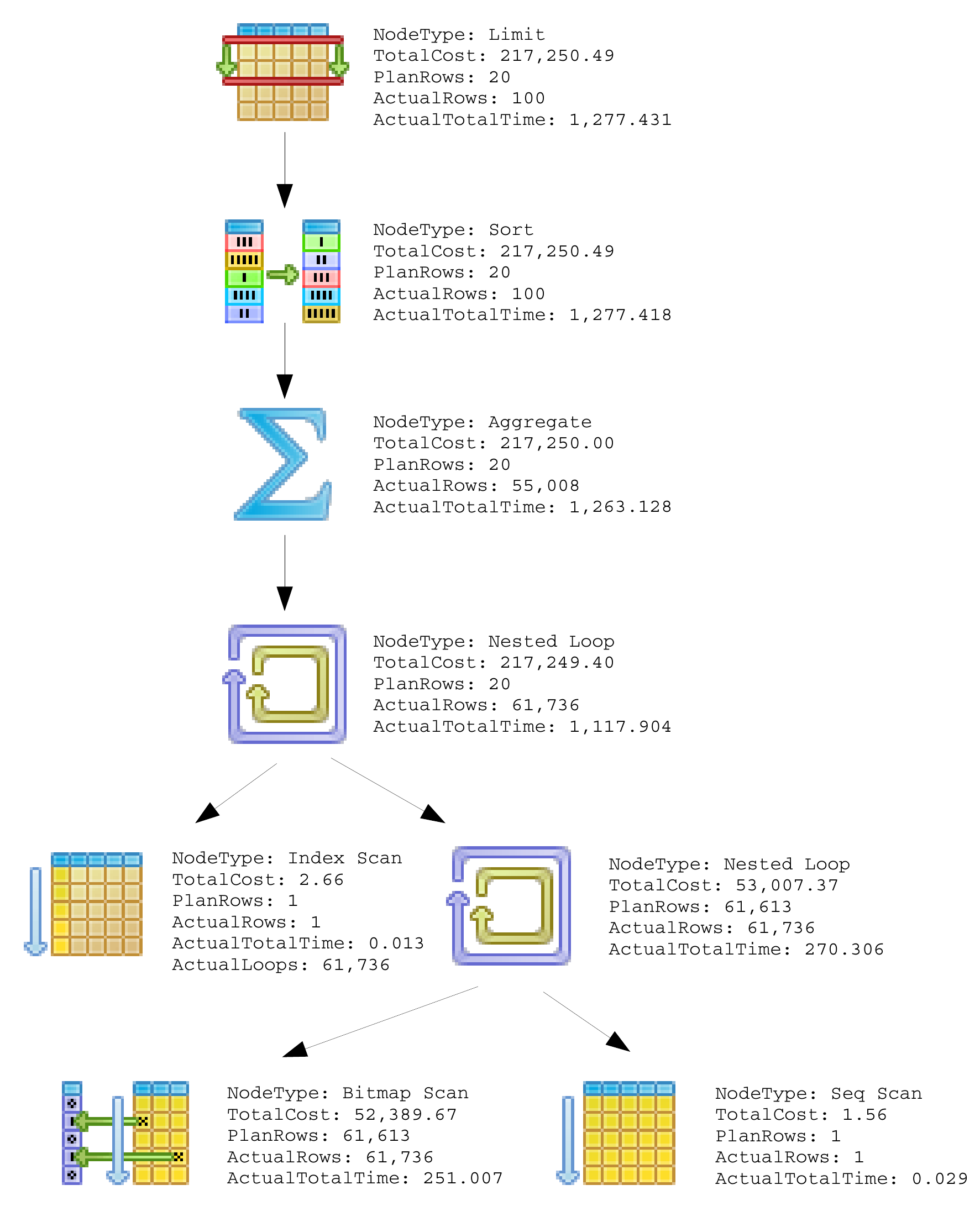}
	\caption{The execution plan for an instance of Template 93 for TPC-DS shows the plan nodes, their cost and execution times.}
	\label{fig:ExamplePlan}
\end{figure}

\subsection{The Optimizer Cost}
\label{sec:OpCost}

During the planning/optimization stage of query execution, the DBMS must consider different join orders and different algorithms for executing some of the operations (e.g. Hash Join, Nested Loop Join or Merge Join). Each set of options results in a different query execution plan. A query plan is a binary tree, where each node represents an operator (such as a Sequential Scan or Sort). The query is executed bottom-up with results from the leaf nodes being fed to their parents. Fig. \ref{fig:ExamplePlan} shows an example of a query plan, for a query derived from template 93 of the TPC-DS benchmark \cite{Poess2007}.

As part of the planning process, the planner/optimizer must rank competing plans in order to select the optimal one. This is done by assigning a cost (the optimizer cost) to a plan based on an analytical model of the resources required by different plans \cite{Selinger1979}.

Taking PostgreSQL as an example, the analytical cost model is given in equation (\ref{eqn:CostModel}), below. The $c$ values are the cost constants that define how expensive different operations are and the $n$ values are the relevant cardinality estimates (the number of rows predicted by the planner to be output by each operator). Each term corresponds to a different cost element, e.g. \textit{seq\_page\_cost} and \textit{cpu\_operator\_cost}. 

\begin{equation}
C_O = n_s\cdot c_s + n_r\cdot c_r + n_t\cdot c_t + n_i\cdot c_i + n_o\cdot c_o
\label{eqn:CostModel}
\end{equation}

The analytical cost model is designed, in theory, to predict how long a query plan would take to execute. However, it is sufficient for the model to correctly rank competing plans such that the cheapest plan is also the fastest, even if the cost is not an accurate prediction. Since this is a somewhat easier task, in practice the optimizer cost is rarely an accurate prediction of execution time. In fact, in PostgreSQL, the optimizer cost is given in arbitrary units and is not a serious attempt at a prediction of execution time. 

It is the inability to use the optimizer cost as a prediction of execution time that motivates research into Query Performance Prediction (QPP). 

\subsection{Related Work}
\label{sec:RelatedWork}

\begin{table*}[t]
	\resizebox{\textwidth}{!}{%
		\centering
		\def\arraystretch{1.1}%
		\begin{tabular}{llllrrc}
			\\ \toprule
			{\bf \multirow{2}{*}{Method}}      & {\bf \multirow{2}{*}{Paper}}            & {\bf \multirow{2}{*}{Level}} & {\bf \multirow{2}{*}{Features}}          & \multicolumn{2}{c}{\bf Relative Error}   & {\bf Training}    \\
			
			{}      & {}   & {} & {}   & {\bf Mean} & {\bf Median}  & {\bf Time}    
			\\ 
			\midrule	
			Nearest Neighbour & \cite{Akdere2012} & Plan                      & Flattened Plan          & 2.1\%  & ---~~~       & ``minutes to hours''     \\ 
			
			Support Vector Regression & \cite{Akdere2012}      & Plan                      & Flattened Plan+  & 6.8\% & ---~~~  & \textgreater 3 seconds   \\ 
			
			Support Vector Regression & \cite{Akdere2012}   & Operator                  & 9 per Operator       & 53.9\% & ---~~~ & \textgreater 3 seconds    \\ 
			
			Regression Trees & \cite{Li2012}   & Operator                  & 9+ per Operator       & 26.0\% & ---~~~ & 2.6 to 36.75 seconds    \\ 
			
			Tuning and Sampling & \cite{Wu2013}       & Operator                  & N/A       & 39.0\%    & ---~~~   &  \textgreater 2 seconds \\ 
			
			
			Nearest Neighbour & 			Our Result       & Plan                      & Optimizer Cost          & 1.5\%   & 0.8\%$^{1}$      & $\approx$ 2 ms                 \\ 
			\bottomrule
		\end{tabular}
	}
	$^{1}$\footnotesize{The median relative error for 10GB TPC-DS was 4.8\%. We list only uniform TPC-H because all papers in the literature used that benchmark and not all used TPC-DS.}
	\caption{Comparison of our main result with the existing literature for the uniform 10GB TPC-H benchmark.}
	\label{tbl:BackgroundQPP}
\end{table*}

The first work on predicting execution time was by Gupta \textit{et al.} who aimed to provide an upper and lower bound for the execution time of a query \cite{Gupta2008}. They used historical data to construct a binary decision tree with a unique classifier at each internal node to direct searches. The leaf nodes contained time ranges in the form $t_l ... t_u$ where $t_l$ and $t_u$ are the lower- and upper-bound on the execution time of the queries that fall in that leaf node.

Following that work, Ganapathi \textit{et al.} proposed using nearest-neighbour regression to provide predictions of resource usage, not just execution time \cite{Ganapathi2009}. Nearest-neighbour regression is a powerful, non-parametric method that requires no training and performs all its work when a prediction is required \cite{Altman1992}. Instead of proposing a model and tuning its parameters, nearest-neighbour regression works by identifying the $k$ most similar data points in the training data and assuming that the target values associated with them will be similar to the value associated with the new data point. 

In this case, this method works by recording a summary of every previously run query along with its execution time. When a new query is being run and a prediction is needed, we find the $k$ most similar previously-run queries and average their execution times to provide a predicted execution time. The underlying principle is that the execution time of similar queries (as determined by the summary of the query plan) will be similar.

To identify the nearest neighbours, Ganapathi \textit{et al.} took a flattened version of the execution plan which contained the number of instances of each operator and the sum of the cardinalities  of those operators. When an operator does not appear in the plan then its entries are set to zero. For example, the features for the plan in Fig. \ref{fig:ExamplePlan} (excluding the zero entries) would be:

\noindent\resizebox{\columnwidth}{!}{%
	\begin{tabular}{|c|c|c|c|c|c|c|c|c|c|c|c|c|c|}
		\hline
		\multicolumn{2}{|c|}{Limit} & \multicolumn{2}{c|}{Sort} & \multicolumn{2}{c|}{Aggregate} & \multicolumn{2}{c|}{Nested Loop} & \multicolumn{2}{c|}{Index Scan} & \multicolumn{2}{c|}{Seq Scan} & \multicolumn{2}{c|}{Bitmap Scan} \\ \hline
		1            & 20           & 1           & 20          & 1             & 20             & 2             & 61633            & 1              & 1              & 1             & 1             & 1             & 61613            \\ \hline
	\end{tabular}
}

Because the feature set is large, they also applied dimensionality reduction which they reported as taking ``minutes to hours''. They found that, using this method, they could predict the execution time of 85\% of TPC-DS (1GB) queries to within 20\% of their true values. Others applying their method found a mean relative error of just 2.1\% for 10GB TPC-H \cite{Akdere2012}.

Akdere \textit{et al.} considered the use of Support Vector Regression \cite{Akdere2012}. Support Vector Regression is the regression version of Support Vector Machines \cite{Smola2004} and is able to model non-linear relationships. Support Vector Machines work by finding a hyperplane that best divides the data points of different categories. The hyperplane becomes a decision boundary and predictions are made based on the location of a new data point in relation to the boundary. Using kernels to project the points into higher dimensions before trying to find the hyperplane allows Support Vector Machines to generate non-linear decision boundaries. 

Aside from considering a new learning algorithm, Akdere \textit{et al.} made the significant contribution of defining plan level and operator level modelling. Plan level modelling is when features are taken from the overall plan and a single prediction is made for the entire plan. For operator level modelling, the chosen query plan is decomposed into its constituent operators and unique models are trained for each operator type. Predictions are then made for each operator in the plan and these predictions are summed to give a final prediction for the entire plan. 

Plan level modelling has the advantage of being able to capture interactions between operators. On the other hand, it is reasonable to assume that the model that is most appropriate for one operator is not the optimal choice for another. For example, the behaviour of Sequential Scans is likely to be very different to that of a Hash Join. Operator level modelling may be better able to capture these differences.

Akdere \textit{et al.} used the same features as Ganapathi \textit{et al.} for their plan level models but added seven further features with information about the overall plan such as its cost and the number of rows that the plan will return. For the operator level models they used nine features per operator including values such as the number of input and output rows and the predicted run times of its child nodes. 

Overall, they found that they could predict the execution time with a mean relative error of 6.75\% for the plan level and 7.30\% for the operator level methods on a 10GB TPC-H benchmark. It should be noted that all their experiments excluded four of the 22 TPC-H templates because they took too long to execute. For the same reason, fewer queries from template 9 were included than from other templates, which is significant because of the much higher variation among template 9 queries that make it harder to predict for these queries. At the operator level, four more templates were excluded because they contained ``PostgreSQL-specific structures ... which lead to non-standard (i.e. non tree-based) execution plans'' \cite{Akdere2012}. 

Li \textit{et al.} considered the problem of applying machine learning solutions to queries that are either running on larger datasets or derived from different templates than those in the training data \cite{Li2012}. They used regression trees to predict execution times at the operator level and then designed and trained ``scaling functions''  that allow the predictions derived from the regression tree to be scaled for queries with cardinalities not previously seen in the training data. 

They used different features for each operator. There were seven features that were common to all the operators such as the number of input and output tuples. Others, such as the size of input tables, applied only to specific operators (scans in this case). They did not include the optimizer cost as one of the features, although for scans they did include the optimizer's estimated I/O cost. Their approach was not as accurate as using nearest-neighbour or SVR, with a mean relative error of 26.0\% for uniform TPC-H. However, they had the advantage of being able to achieve a similar error even if the data size was changed.

More recently, Wu \textit{et al.} proposed using a so-called white-box approach instead of machine learning \cite{Wu2013}. They argued that even if the optimizer cost could not be used to predict the execution time that was because the parameters used in the calculation of the cost were incorrect, rather than a flaw in the analytical model itself. They therefore proposed a two-step process to use the analytical model to produce optimizer costs that are accurate predictions. The first step is to tune the cost constants to make them more accurate for the hardware of the system. This was done through a series of carefully constructed queries. Then, more accurate cardinality estimates are made for the chosen query plan using increased sampling so that the inputs to the formula are more accurate. 

The overhead of the sampling was reported as being between 4\% and 20\% of total execution time which is not insignificant. Moreover, the method is not as accurate as the post-processing approaches, achieving a mean relative error of 39.0\% for uniform TPC-H.

Table \ref{tbl:BackgroundQPP} compares our main result with the results in the existing literature.

\section{Experimental Setup}
\label{sec:Setup}

All of the queries in our work were executed with PostgreSQL v9.4.4 on a machine which runs CentOS 6.5 and has 2 sockets and a 6-core Intel Xeon E5-2620 processor per socket, as well as 32GB memory. 

The machine learning algorithms were run off-line using scikit-learn for Python \cite{Pedregosa2011} on a standard desktop. We adopted the k-fold cross validation experimental method taking $k=5$ \cite{Refaeilzadeh2009}. In this method, all of our collected data is used as both training and testing data but not at the same time. Specifically, the data is divided into $k$ independent chunks or folds and each fold serves as the testing data for one iteration and as part of the training data for $k-1$ iterations. When it is the testing data then the model is tuned only from the remaining data. This is done to prevent \textit{overfitting} which occurs when the model learns how to accurately predict the training data but has very limited ability to extrapolate correctly to examples outside of the training data. 

\subsection{Benchmarks}
\label{sec:Benchmarks}

We use three standard decision support benchmarks which have been used in the literature. All three capture the decision support system behaviour by providing query templates. Each template includes one or more variables whose values can be changed to produce different query instances. Every instance derived from the same template performs a similar analytical task.

The first benchmark is TPC-H, which consists of 22 templates \cite{Council2008}. The data in TPC-H is uniformly distributed meaning that the number of rows in a given column that match a given value is approximately constant, regardless of the value chosen. This means that we would expect different query instances from the same template to be very similar in their plans and execution times because the choice of constant values in the query has little effect on the number of rows returned.

We therefore also use a benchmark generated from the TPC-H Skew tool with a skew of Z=1 \cite{Chaudhuri} \footnote{We used data generated by the tool, available from \url{http://www.cs.toronto.edu/~consens/tab/} because the tool itself was no longer available}. The data in this benchmark is compatible with the TPC-H queries but follows a Zipfian (power-law) distribution. This means that the number of rows in a given column that match a given value can vary significantly with different choices of that value. This should lead to more variation in plans and execution times among queries derived from the same template.

All of our experiments with these benchmarks are with 10GB versions and use 100 instances of each template (a total of 2,200 queries). Each query was run in isolation and the chosen query plan and the execution time were recorded.

To confirm the differences in these two benchmarks we compared the Coefficient of Variation (CoV) of the execution times within each template for each of the benchmarks. The CoV is the standard deviation divided by the mean and provides a normalised view of the spread of values. 

\begin{figure}[t]
	\centering
	\includegraphics[width=\columnwidth]{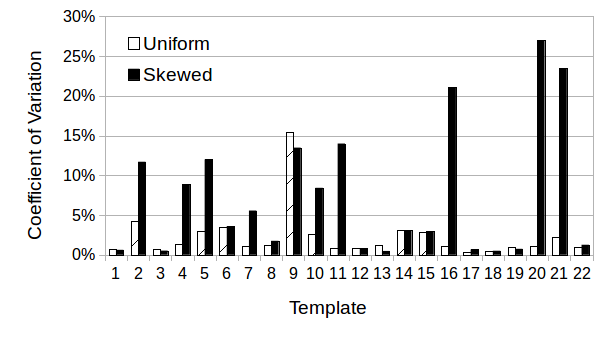}
	\caption{The Coefficient of Variation of execution times within each template in the TPC-H benchmark for both the uniform and skewed versions.}
	\label{fig:CoV_TPCH}
\end{figure}

Fig. \ref{fig:CoV_TPCH} shows the CoV for both TPC-H benchmarks. As expected, in most cases the variation for the Uniform version of TPC-H is very small and the CoV for the Skewed version is often significantly higher. Template 9 stands out as an exception for the Uniform version. It turns out that this is not because of the underlying data but because of incorrect assumptions in the planner. 

One line of the template is ``\texttt{p\_name like `\%:1\%'}'' where different query instances insert different character strings from a pool of choices in place of \texttt{:1} . In the 10GB TPC-H Uniform benchmark, the number of rows matching that condition is approximately 108,500 regardless of the chosen string. In some cases the planner estimates the number of returned rows to be approximately 160,000 and in other cases it predicts that around 80,000 rows will be returned. This difference results in the planner sometimes choosing to perform an aggregate before a sort (when estimating 160,000) and sometimes the other way around (when estimating 80,000). When the aggregate is performed first the queries take approximately 150 seconds, whereas performing the sort first results in queries taking approximately 200 seconds. This is why template 9 shows a larger CoV even for the Uniform version.

\begin{figure}[t]
	\centering
	\includegraphics[width=\columnwidth]{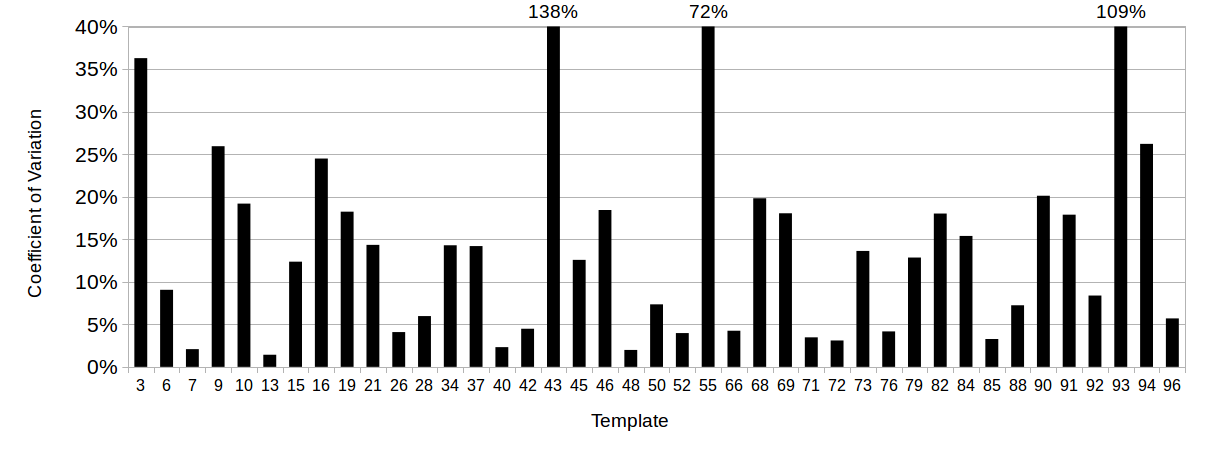}
	\caption{The Coefficient of Variation of execution times within each template in the TPC-DS benchmark.}
	\label{fig:CoV_TPCDS}
\end{figure}

The third benchmark is the more modern TPC-DS \cite{Poess2007} which has 99 query templates although only 41 are compatible with PostgreSQL\footnote{\url{http://blog.pgaddict.com/posts/performance-since-postgresql-7-4-to-9-4-tpc-ds}}. As with TPC-H, we used a 10GB benchmark and generated 100 queries per template. TPC-DS is a more complex benchmark and has more skew as can be seen from the much higher CoV values in Fig. \ref{fig:CoV_TPCDS}.

\subsection{Metrics}
\label{sec:Metrics}

To compare our results with those published using other methods, we utilise the same metrics as in previous work. The primary metric is the relative error, as shown in equation (\ref{eqn:MRE}). This gives a relative measure of the difference between the predicted and actual target values \cite{Akdere2012}. Akdere \textit{et al.} reported the mean relative error in their work and we do so too. However, we also report the median because the relative error can include significant outliers which skew the mean.

\begin{equation}
Relative~Error = \displaystyle\frac{|actual - predicted|}{actual}
\label{eqn:MRE}
\end{equation}

A second metric we use throughout is the proportion of queries that were correctly predicted to within 20\% of the true execution time. This metric was used by Ganapathi \textit{et al.}, although they did not justify the choice of 20\%. Nevertheless, it does provide a useful high-level view of the performance of the methods and allows direct comparison to their work. This metric is also important because the mean and median can hide the spread.

\section{Evidence Against Post-Processing}
\label{sec:EvidenceAgainst}
In this section, we highlight the limited nature of the evidence against post-processing in the literature. Despite the strong conclusion of Wu \textit{et al.} and the consensus, we found only three works with results concerning the use of the optimizer cost for predicting execution times \cite{Ganapathi2009,Li2012,Akdere2012} and all are concerned with linear regression at the plan level.

Ganapathi \textit{et al.} \cite{Ganapathi2009} attempted to draw a \textit{line of best fit} between the costs from HP's Neoview optimizer and the execution times for 1GB TPC-H. They found that the cost was up to 100 times away from the best fit line. Li \textit{et al.} \cite{Li2012} also drew a \textit{line of best fit} using results from Microsoft's SQL Server on TPC-H Skewed of sizes between 1GB and 10GB. They first removed any plans with cardinality estimates that were more than 10\% out to try and improve the likelihood of success. They concluded, however, that even with this filtering ``there are significant differences between the estimated CPU cost and real CPU time for many queries''.

The only quantitative result we could find in the literature was from Akdere \textit{et al.} who trained a \textit{linear} regression model at the plan level \cite{Akdere2012}. They tested the resulting model on I/O intensive queries from TPC-H using PostgreSQL and found a mean relative error of 120\% with a minimum of 30\% and a maximum of 1744\%. As we noted earlier, the mean is not the best metric in this case because of the high-skew in the errors.

\begin{table}[t]
	\resizebox{\columnwidth}{!}{%
		\begin{tabular}{lrrr}
			\toprule
			{\textbf{\multirow{2}{*}{Linear Regression}}}  & \multicolumn{2}{c}{\textbf{TPC-H}} & \textbf{TPC-DS} \\ 
			{}  & Uniform          & Skewed           &                 \\ 
			\midrule
			\textbf{Plan Level} & & & \\
			Mean Relative Error & 100.4\% & 119.1\% & 352,631.0\% \\
			Median Relative Error & 42.3\% & 50.9\% & 75.0\% \\ 
			{Queries with $<$20\% Error} & 15.9\% & 13.6\% & 14.8\%\\
			\bottomrule
		\end{tabular}
	}
	\caption{The relative error results of linear regression using only the optimizer cost at the plan level.}
	\label{tbl:LinReg}
\end{table}

\begin{figure}[t]
	\centering
	\includegraphics[width=\columnwidth]{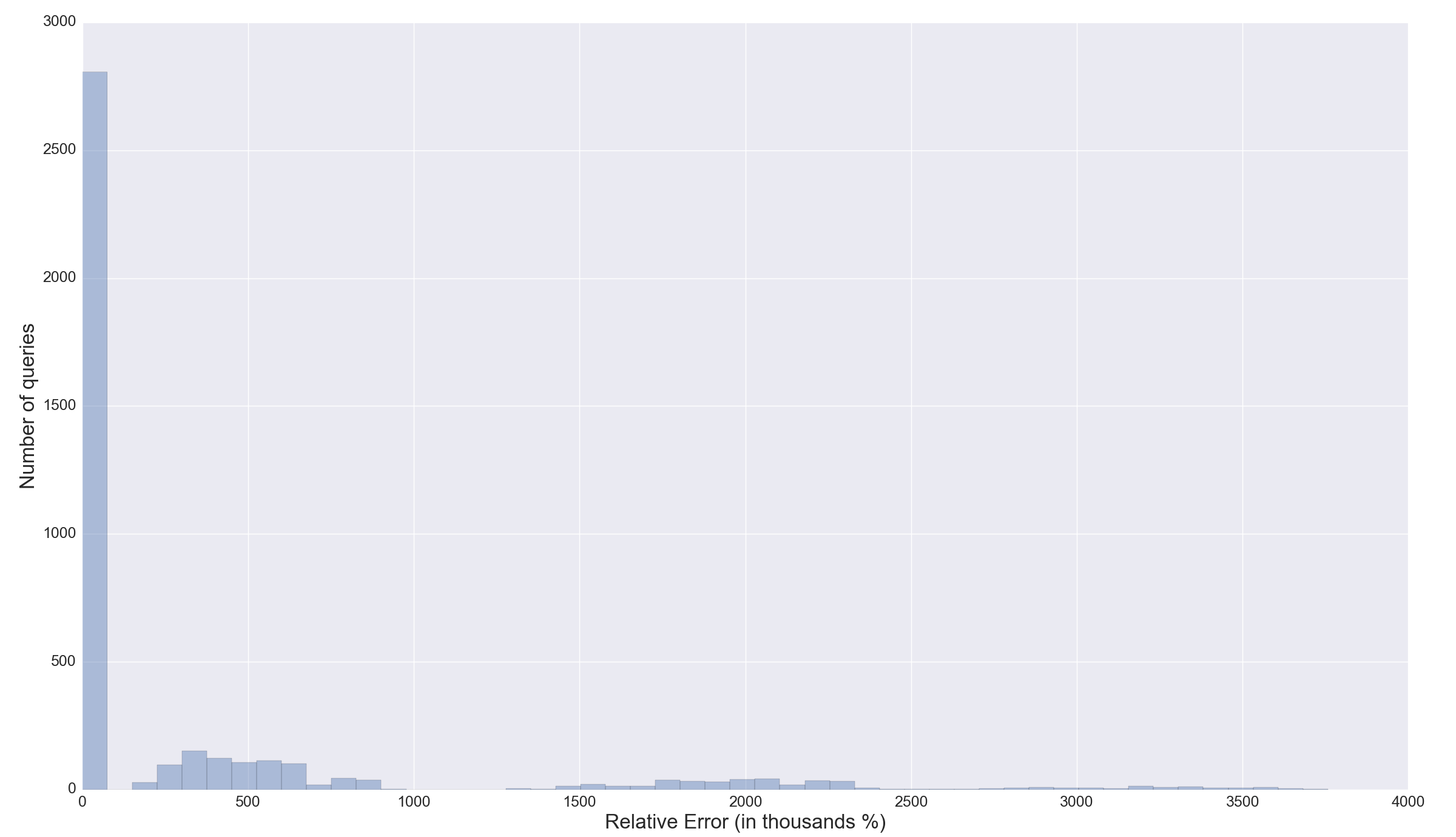}
	\caption{The long tail distribution of the relative errors shows the importance of using the median as the typical value rather than the mean}
	\label{fig:ErrorDistribution}
\end{figure}

For the sake of completeness we repeated these experiments for the three benchmarks. We found a mean relative error of 100.4\% for the uniform 10GB TPC-H benchmark which is similar to the 120\% reported by Akdere \textit{et al.}, although we found a minimum of 3.3\% and a maximum relative error of 500.1\% which are both lower than they found.

Our results are summarised in Table \ref{tbl:LinReg}. They show that the optimizer cost cannot be used as the sole feature with linear regression at the plan level. 

Figure \ref{fig:ErrorDistribution} shows the distribution of the relative errors for TPC-DS and shows a long tail distribution. The long tail distribution is encountered for all methods that we consider and for all benchmarks, although it is more significant for TPC-DS than the two TPC-H benchmarks. With such a distribution of errors, focussing on the mean relative error can be misleading and we prefer to focus on the median relative error which gives a more accurate representation of the typical error. However, we include the mean relative error in all cases because that is the metric used in the literature. The long tail distribution explains the extremely large mean relative error for TPC-DS shown in Table \ref{tbl:LinReg}.

\section{Post-Processing the Optimizer Cost}

In the previous section, we confirmed that using linear regression at the plan level does not lead to accurate predictions of query execution time. In this section, we apply more complex and non-linear approaches. Specifically, we first apply linear regression at the operator level, then apply a power-law regression at both plan and operator levels. Finally, we apply SVR at both plan and operator levels.

\subsection{Operator-Level Linear Regression}
\label{sec:OpLevelLinReg}

\begin{table}[t]
	\resizebox{\columnwidth}{!}{%
		\begin{tabular}{lrrr}
			\toprule
			{\textbf{\multirow{2}{*}{Linear Regression}}}  & \multicolumn{2}{c}{\textbf{TPC-H}} & \textbf{TPC-DS} \\ 
			{}  & Uniform          & Skewed           &                 \\ 
			\toprule
			\textbf{Operator Level} & & & \\
			Mean Relative Error & 52.6\% & 60.0\% & 243,885.5\%\\
			Median Relative Error & 19.9\% & 21.3\% & 65.2\%  \\ 
			{Queries with $<$20\% Error} & 50.4\% & 45.7\% & 11.3\%\\
			\bottomrule
		\end{tabular}
	}
	\caption{The relative error results of Linear Regression using only the optimizer cost at the operator level.}
	\label{tbl:OpLevelLinReg}
\end{table}

As discussed in Section \ref{sec:RelatedWork}, Akdere \textit{et al.} introduced the distinction between plan-level and operator-level modelling. Plan-level modelling is when a single model is created for the entire query execution plan. For linear regression this means taking the optimizer cost of the overall plan and fitting it into a linear regression model. We have already seen that this is not effective.

Operator-level modelling works by creating unique models for each operator type. A query plan is then broken down into its component operators and a separate prediction is made for each operator. These predictions are then summed to give a prediction of total execution time. This is similar to the way the optimizer itself works because the query optimizer calculates a cost for each operator and the total cost is the sum of all the individual costs.

The advantage of operator-level modelling is that it allows more flexibility. It stands to reason that the model that is more appropriate for a Sequential Scan, for example, is not appropriate for a Sort. The cost is that we have to perform more training.

Table \ref{tbl:OpLevelLinReg} shows that applying linear regression at the operator level approximately halves the median relative error for the TPC-H benchmarks and more than doubles the proportion of queries that are predicted to within 20\% of their true run-times. For TPC-H Uniform, the median relative error is now below 20\% and for the Skewed version it is just over, at 21.3\%. The proportion of queries that can be predicted to within 20\% of their true values is 50.4\% for the Uniform TPC-H and 45.7\% for the Skewed version.

For TPC-DS the median error also falls, from 75.0\% to 65.2\%, but the proportion with less than 20\% errors actually falls from 14.8\% to 11.3\%. That suggest that for the TPC-DS benchmark, the operator level models are being overfit and therefore do not generalise well to the test data.

\subsection{Power-Law Regression}
\label{sec:PowerLaw}

\begin{figure}[t]
	\centering
	\begin{minipage}{0.33\textwidth}
		\centering
		\includegraphics[width=0.9\textwidth]{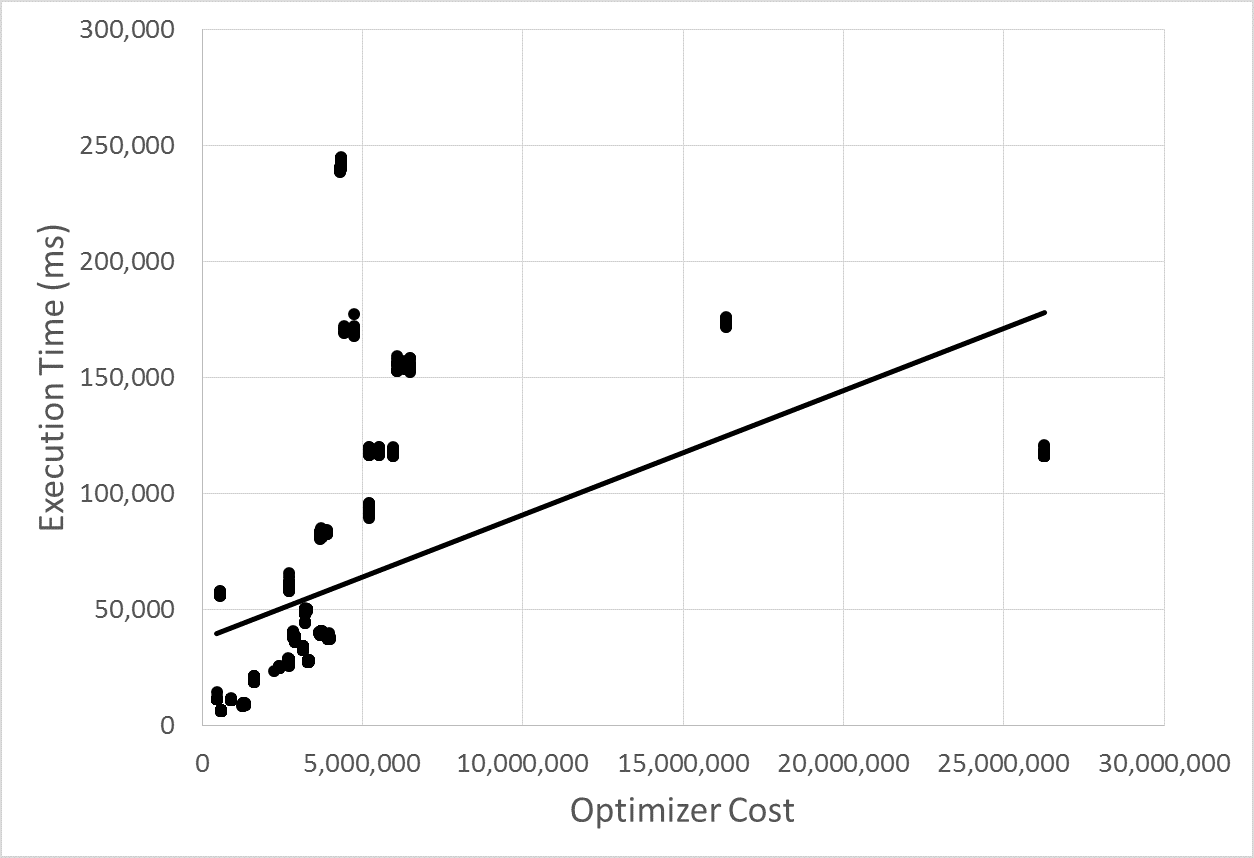}\\
		(a) TPC-H Uniform
	\end{minipage}\hfill
	\begin{minipage}{0.33\textwidth}
		\centering
		\includegraphics[width=0.9\textwidth]{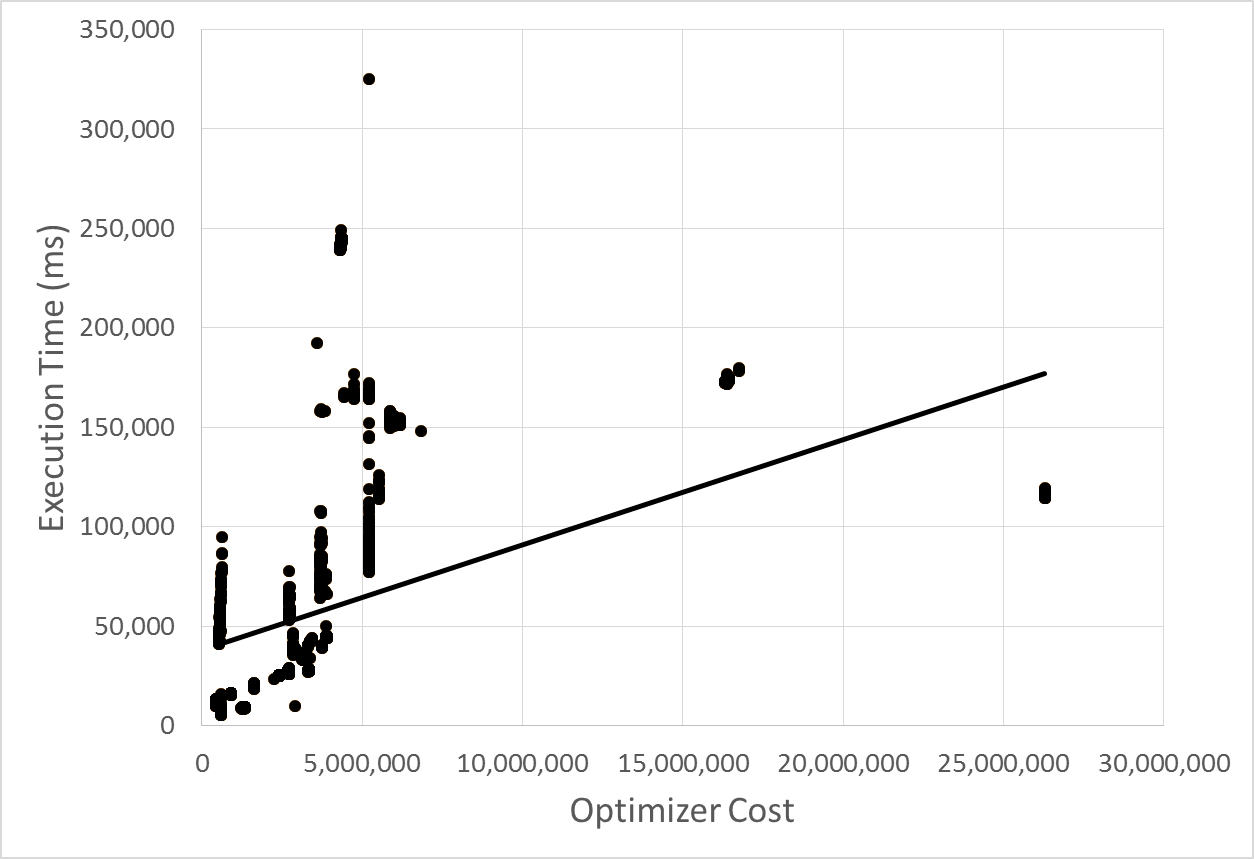}\\
		(b) TPC-H Skew
	\end{minipage}
	\begin{minipage}{0.33\textwidth}
		\centering
		\includegraphics[width=0.9\textwidth]{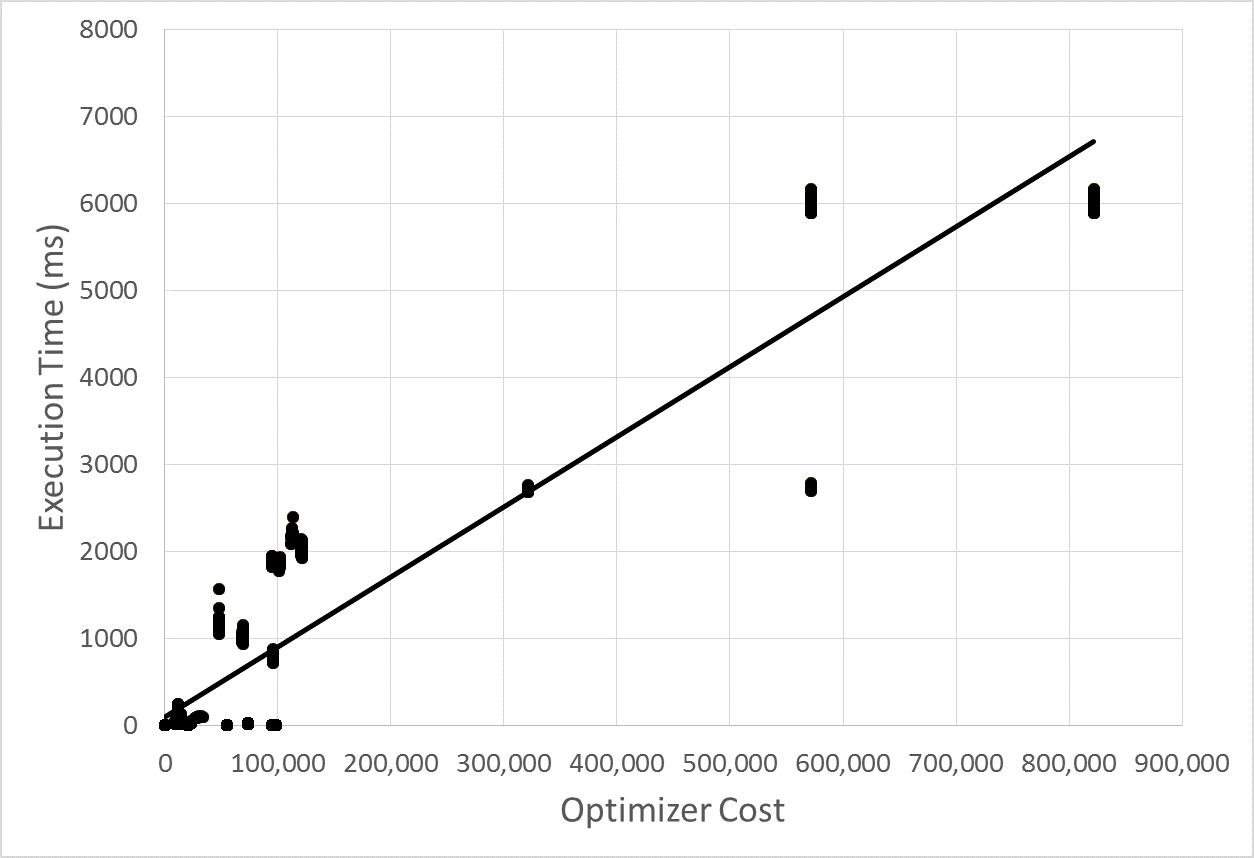}\\
		(c) TPC-DS
	\end{minipage}
	\caption{Plotting optimizer cost against execution time shows a lack of linear correlation, indicating why linear regression applied to the optimizer cost provides a poor prediction of execution time.}
	\label{fig:LinearRegression}
\end{figure}

\begin{figure}[t]
	\centering
	\begin{minipage}{0.33\textwidth}
		\centering
		\includegraphics[width=0.9\textwidth]{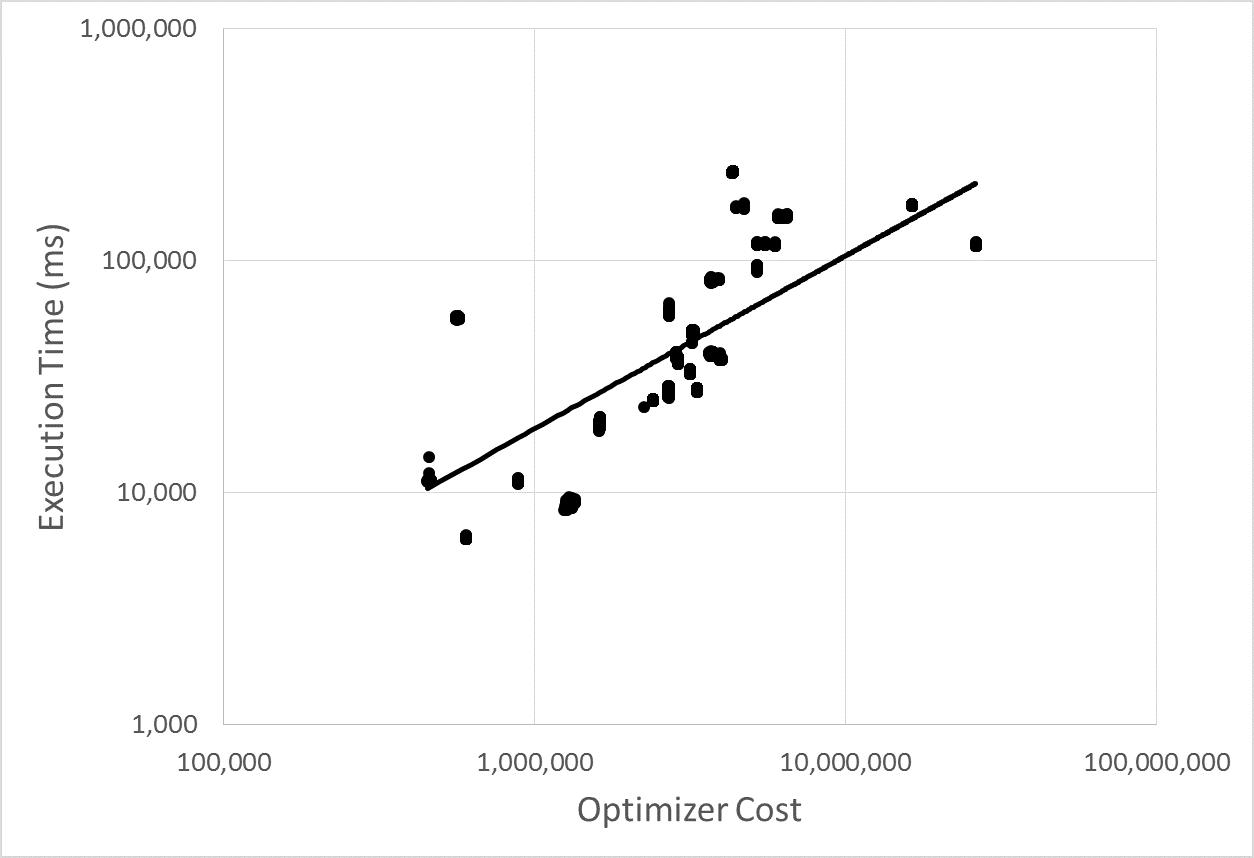}\\
		(a) TPC-H Uniform
	\end{minipage}\hfill
	\begin{minipage}{0.33\textwidth}
		\centering
		\includegraphics[width=0.9\textwidth]{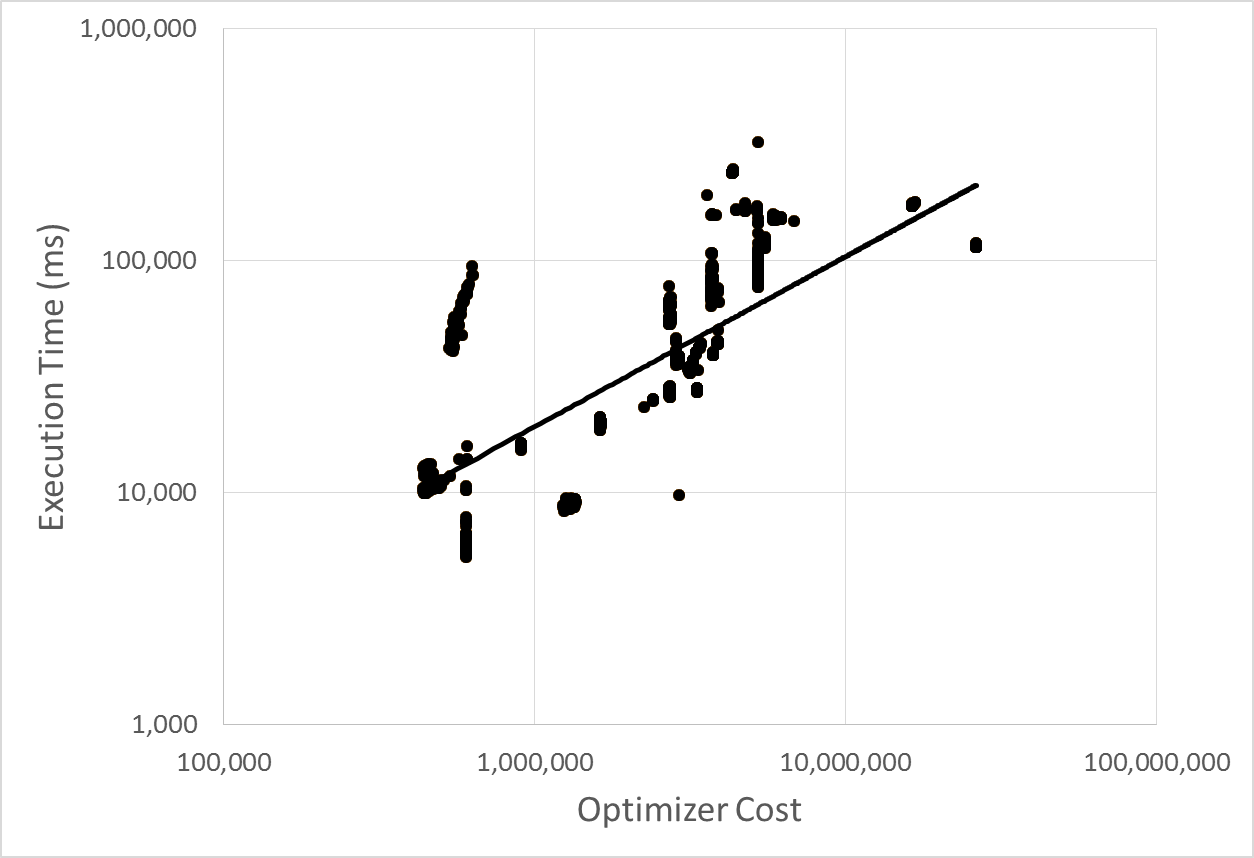}\\
		(b) TPC-H Skew
	\end{minipage}
	\begin{minipage}{0.33\textwidth}
		\centering
		\includegraphics[width=0.9\textwidth]{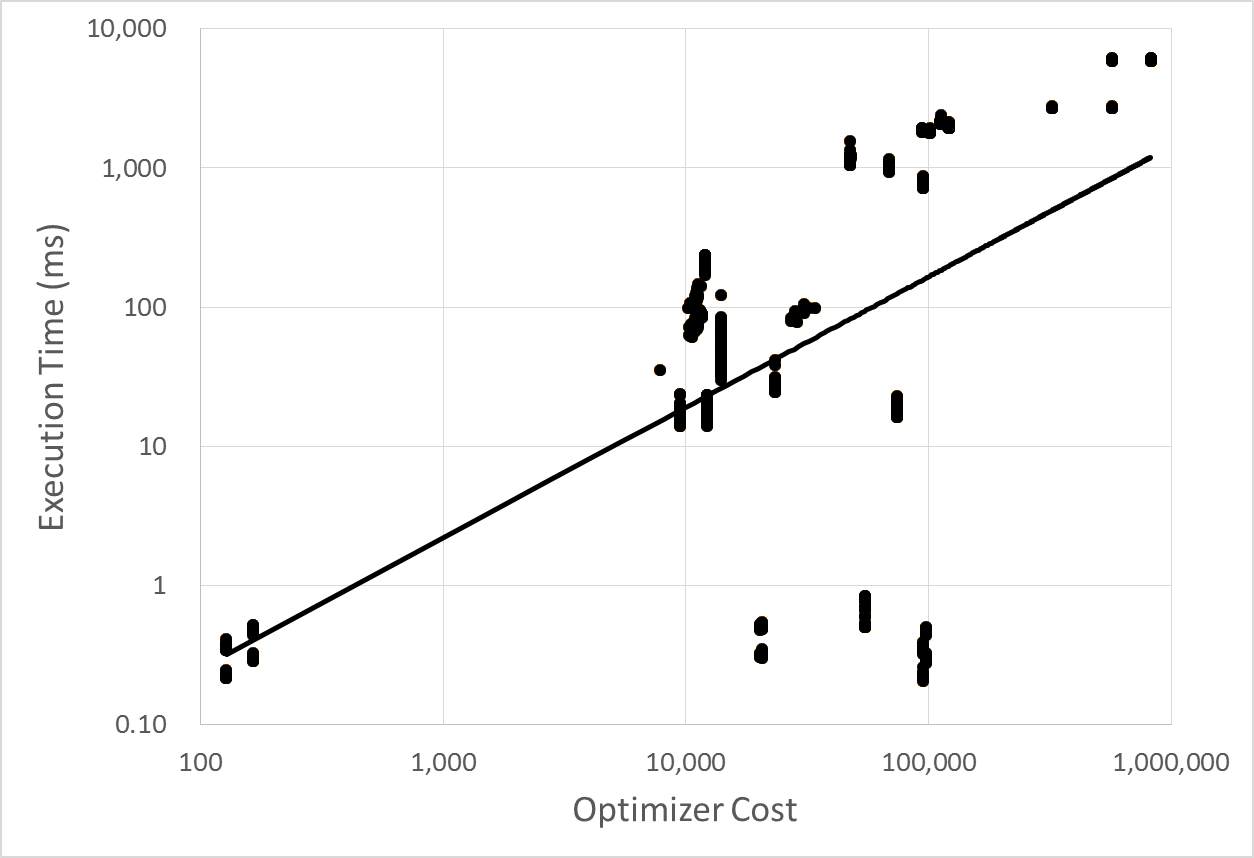}\\
		(c) TPC-DS
	\end{minipage}
	\caption{Plotting optimizer cost against execution time on a log-log scale suggests a power-law distribution because a linear correlation appears.}
	\label{fig:PowerLaw}
\end{figure}

Figure \ref{fig:LinearRegression}(a)-(c) shows the relationship between the optimizer cost and execution time of the queries in the three benchmarks. It shows that there is no linear correlation between the two and this explains why linear regression at the plan level produces poor predictions. However, Figure \ref{fig:PowerLaw}(a)-(c) shows the same relationship but plotted on a log-log scale. On this scale, at least for the two TPC-H benchmarks, a moderate linear correlation appears with $R^2$ values of 0.57 and 0.56 for the uniform and skewed versions. For TPC-DS the $R^2$ value is 0.31 which, although lower, still indicates a weak linear correlation \cite{Moore2013}.

We therefore consider a simple regression model based on fitting a power-law correlation between the optimizer cost and the execution time. For completeness, we also consider this at the operator level. To model this relationship we can use the ordinary least squares algorithm in the same way as for linear regression but we first take the logarithms of the optimizer cost and execution time. A linear regression model of the logarithms is the same as a power-law regression of the original data.

\begin{table}[t]
	\resizebox{\columnwidth}{!}{%
		\begin{tabular}{lrrr}
			\toprule
			{\textbf{\multirow{2}{*}{Power-Law Regression}}}  & \multicolumn{2}{c}{\textbf{TPC-H}} & \textbf{TPC-DS} \\ 
			{}  & Uniform          & Skewed           &                 \\ 
			\toprule
			\textbf{Plan Level} & & & \\
			Mean Relative Error & 49.7\% & 59.5\% & 34,800.3\%\\
			Median Relative Error & 40.2\% & 43.5\% & 92.5\%  \\ 
			{Queries with $<$20\% Error} & 25.6\% & 22.6\% & 0.8\%\\
			\midrule
			\textbf{Operator Level} & & & \\
			Mean Relative Error & 3,367.4\% & 3,780.3\% & 1.6E+15\%\\
			Median Relative Error & 3,174.5\% & 3,427.9\% & 2,647.0\%  \\ 
			{Queries with $<$20\% Error} & 0.0\% & 0.0\% & 0.0\%\\			
			\bottomrule
		\end{tabular}
	}
	\caption{The relative error results of Power-Law Regression using only the optimizer cost at the plan and operator levels.}
	\label{tbl:PowerReg}
\end{table}

Table \ref{tbl:PowerReg} shows the results for both the plan level and operator level. At the plan level, compared to plan-level linear regression, applying the power-law leads to a significant improvement for the TPC-H benchmarks. The median relative error falls moderately, but the proportion of queries with errors of less than 20\% increases from 15.9\% with linear regression (TPC-H Uniform) to 25.6\%. This is mainly driven by a decrease in larger errors which is also shown in the large decrease in mean relative error from 100.4\% and 119.1\%, in the uniform and skewed versions respectively, to 49.7\% and 59.5\%. 

For TPC-DS, we also observe a very large reduction in mean relative error, down to 34,800\% from 352,631\%. Of course, this number is still extremely large and so the decrease is of little benefit. Moreover, median relative error is increased and the proportion of queries with small errors falls greatly. Results for the operator level are even worse with enormous errors for all benchmarks. 

What these results indicate is that when the power-law is an appropriate fit it can lead to reasonably good predictions. Where it is an inappropriate model, however, the errors that are observed are much larger than for linear regression. This is to be expected because with a power-law, small changes in the optimizer cost lead to much larger changes in the predicted execution time than they do for a linear model. Therefore, where the distribution is not, actually, a power-law the errors are exponentially larger.

\subsection{Non-Linear Regression}
\label{sec:SVR}

Here we consider non-linear regression by using Support Vector Regression (SVR) with non-linear kernels.

The basic SVR algorithm can only train linear regression models but using the ``kernel trick'' we can project the original set of features into higher dimensions and find linear correlations in that space. Those linear correlations are equivalent to non-linear correlations in the original feature space. We consider three kernels - linear, polynomial and the Gaussian-based Radial Basis Function \cite{Vert2004}. All three are supported by the widely used LIBSVM package\cite{Chang2011}. SVR was used successfully by Akdere \textit{et al.} for both plan level and operator level prediction with a large feature set \cite{Akdere2012}. 

\begin{table}[p!]
	\resizebox{\columnwidth}{!}{%
		\begin{tabular}{lrrr}
			\toprule
			\multicolumn{1}{l}{\multirow{2}{*}{\begin{tabular}[c]{@{}l@{}}\textbf{Support Vector}\\ \textbf{Regression}\end{tabular}} 
			} & \multicolumn{2}{c}{\textbf{TPC-H}} & \textbf{TPC-DS} \\ 
			
			\multicolumn{1}{l}{} & Uniform          & Skewed           &       {}          \\ 
			\midrule
			\multicolumn{1}{c}{}\textbf{Linear Kernel} & & & \\
			\textbf{Plan Level} & & & \\
			Mean Relative Error & 71.6\% & 73.9\% & 96,693.5\% \\
			Median Relative Error  & 44.2\%   & 43.0\%  & 95.1\%      \\ 
			Queries with $<$20\% Error & 20.4\% & 24.0\% & 11.9\%\\
			\midrule
			\textbf{Operator Level} & & & \\
			Mean Relative Error & 34.2\% & 36.1\% & 19,065.9\%\\
			Median Relative Error  & 30.2\%  & 28.5\% & 76.9\%        \\
			Queries with $<$20\% Error & 35.1\% & 32.2\% & 2.1\%\\
			\bottomrule
			\toprule

			\multicolumn{1}{c}{}\textbf{Polynomial Kernel} & & & \\
			\textbf{Plan Level} & & & \\
			Mean Relative Error & 84.6\% & 95.1\% & 92,890.2\% \\
			Median Relative Error  & 40.7\%   & 47.0\%  & 93.2\%      \\ 
			Queries with $<$20\% Error & 26.0\% & 29.4\% & 15.7\%\\
			\midrule
			\textbf{Operator Level} & & & \\
			Mean Relative Error & 52.1\% & 53.6\% & 17,959.9\%\\
			Median Relative Error  & 56.6\%  & 62.3\% & 94.5\%        \\
			Queries with $<$20\% Error & 18.2\% & 21.7\% & 6.0\%\\
			\bottomrule
			\toprule
						
			\multicolumn{1}{c}{}\textbf{RBF Kernel} & & & \\
			\textbf{Plan Level} & & & \\
			Mean Relative Error & 49.1\% & 54.7\% & 97,475.9\% \\
			Median Relative Error  & 32.3\%   & 40.9\%  & 96.6\%      \\ 
			Queries with $<$20\% Error & 43.3\% & 36.5\% & 13.3\%\\
			\midrule
			\textbf{Operator Level} & & & \\
			Mean Relative Error & 28.4\% & 27.0\% & 18,532.1\%\\
			Median Relative Error  & 18.3\%  & 14.9\% & 80.0\%        \\
			Queries with $<$20\% Error & 52.0\% & 61.4\% & 2.8\%\\
			\bottomrule
		\end{tabular}
	}
	\caption{The relative error results of Support Vector Regression, with various kernels, using only the optimizer cost.}
	\label{tbl:SVRReg}
\end{table}

We first consider plan level prediction. Table \ref{tbl:SVRReg} shows the relative error results for the three benchmarks at the plan and operator levels. We focus on the results using the RBF kernel because the RBF kernel, which is recommended as the first choice approach by the LIBSVM developers \cite{Hsu2003}, gives the best results in our experiments.

The results for the TPC-H benchmarks show that, at the operator level, SVR using only the optimizer cost gives better results than those found by Akdere \textit{et al.} using many more features. They reported a mean relative error, at the operator level, of 53.9\% for TPC-H Uniform whereas using only the optimizer cost gives a mean relative error of 28.4\%. In fact, this result is also better than the result reported by Wu \textit{et al.} of 39.0\% and comparable to those found by Li \textit{et al.} of 26.0\%. 

By contrast, for the TPC-DS benchmark, the results with SVR are slightly worse than using simple linear regression with a small drop in the median relative error and the proportion of queries with errors of less than 20\%. 

The results in this section have shown that by using more complex methods, such as SVR at the operator level, post-processing the optimizer cost can compete with some of the published results but with a far smaller set of features. For example, the regression tree method of Li \textit{et al.} achieved a mean relative error of 26.0\% \cite{Li2012} and the pre-processing method of Wu \textit{et al.} achieved a mean relative error of 39.0\% (even with the large overhead of sampling)\cite{Wu2013}.

On the other hand, the accuracy is worse than the best techniques which can achieve mean relative error rates under 10\%. Overall, it is reasonable to conclude that using only the optimizer cost in a predictive model is competitive but not the most effective.

\section{Post-Processing with Nearest-Neighbour Regression}
\label{sec:KNN}


In this section, we consider nearest-neighbour regression using the optimizer cost. The nearest neighbour algorithm is a powerful non-parametric method that can be used for both classification and regression \cite{Altman1992}. It has proven to be the most accurate solution in the literature \cite{Ganapathi2009,Akdere2012}. Akdere \textit{et al.} found a mean relative error of 2.1\% for 10GB TPC-H compared to 6.75\% using their own method based on Support Vector Regression.

Nearest-neighbour regression assumes that there is a pool of data - the neighbours - and that instances that share many characteristics also share the same target value. For example, if we are using nearest-neighbour regression to predict house prices, we would have a collection of data regarding houses (e.g. their size, location, number of bedrooms etc) along with the price of each of those houses. When trying to predict the price of a new house we would identify the $k$ houses from our saved pool which are the closest match to the new house, in terms of the features we have recorded, and predict that the price of the new house is the average of the prices of the $k$ houses we identified as being most similar.

In the context of QPP, nearest-neighbour regression works by assuming that we have a pool of information about previously executed queries and that any new query will be similar to some of those in the pool. The results in the literature show that, where this assumption holds such as with the TPC benchmarks, nearest-neighbour regression can give accurate predictions.

Nearest-neighbour regression is essentially making use of a lookup table and the features are the keys to that table. This is why it may be that although there is no mathematical relationship between the optimizer cost and execution time, nevertheless, the cost can be used to provide accurate predictions via the lookup table.

The challenge for applying nearest-neighbour regression to QPP is to identify a set of features that describe a query plan with enough detail to ensure that the $k$ nearest neighbours identified are indeed those with similar execution times. Ganapathi \textit{et al.} did this by using a flattened version of the query plan (see Section \ref{sec:RelatedWork}). In this section, we show that the optimizer cost alone is sufficient, with no loss of accuracy.

We note that the analytical model from which the optimizer cost is generated is non-trivial (see Section \ref{sec:OpCost}) and includes cardinality estimates which may vary significantly between queries. Moreover, the plan-level optimizer cost is the sum of the costs of each operator and the templates in the TPC benchmarks are non-trivial meaning that they contain more than just one or two operators. It is therefore reasonable to suppose that, serendipitously, the optimizer costs of two queries will only be similar if the two queries have essentially the same plan, perhaps varying only a small amount in their cardinality estimates because of differing predicates in one or two operators. If this is the case, then nearest-neighbour regression using only the optimizer cost would be just as accurate as using the flattened version of the query plan.

We also note that cardinality estimation errors will have limited impact on the effectiveness of our approach. Suppose that a given predicate applied to a base relation has selectivity $s_1$ but the planner estimates it to have selectivity $s_2$ which is significantly different. This will certainly affect the optimizer cost. However, if two queries both contain the same predicate applied to the same base relation then the cardinality estimation error is the same in both queries and so their optimizer costs are both affected in the same way and to the same extent. As far as the nearest-neighbour regression algorithm is concerned they may remain the nearest neighbours. This is important because cardinality estimation remains an open-problem \cite{Lohman2014}.

There is an exception to this, however, which is when the cardinality estimation error makes two different queries look the same. This can happen when two different predicates have different selectivities but the planner estimates them to have the same selectivity. In this case the two query plans look very similar when, in actuality when executed, they are not. We note, however, that this problem would affect the flattened plan features used by Ganapathi \textit{et al.} as well. Since the problem is that two different query plans are made to look the same, they look the same whether looking at the optimizer cost or the flattened plan.

By using only the optimizer cost to compare plans, we have three advantages. The first is that our pool of saved data is much smaller. For each previously run query we would need to store only the optimizer cost and the execution time, perhaps 128 bits. The flattened plan requires us to store two pieces of information per operator type per plan as well as the execution time. In our experiments there were a total of 14 operators and so the flattened plan would require storing at least 14 times as much data.

The second advantage is that the search time is greatly reduced. Nearest-neighbour regression typically identifies closest neighbours by finding those neighbours with the smallest Euclidean distance. As the number of features (dimensions) grows, it takes longer to find the nearest neighbours. For this reason, Ganapathi \textit{et al.} used dimensionality reduction as a pre-processing step before using nearest-neighbour regression. But this adds overhead and the dimensionality reduction process can take a very long time (``minutes to hours'' as Ganapathi \textit{et al.} reported). In contrast, with only one feature (the optimizer cost) the time required to find the nearest neighbours is much smaller.

A third advantage is that the optimizer cost is available immediately along with the query plan with no additional overhead. By contrast, to produce the set of features used by Ganapathi \textit{et al.}, the query plan has to be parsed to extract the names of the operators being used and count the number of times they appear. The more complex the query plan, the longer this process will take.

\begin{table}[t]
	\centering
	\resizebox{\columnwidth}{!}{%
		\begin{tabular}{lrrr}
			\toprule
			\multirow{2}{*}{\begin{tabular}[c]{@{}l@{}}\textbf{Nearest Neighbour}\\\textbf{Regression}\end{tabular}} & \multicolumn{2}{c}{\textbf{TPC-H}} & \textbf{TPC-DS} \\
			& Uniform          & Skewed           &                 \\
			\midrule
			\textbf{Plan Level} & & & \\
			Mean Relative Error & 1.7\% & 5.3\% & 3,243.1\%\\
			Median Relative Error  & 0.8\% & 1.1\% & 4.8\% \\
			Queries with $<$20\% Error & 99.5\% & 94.8\% & 78.3\%\\
			\toprule
			\textbf{Operator Level} & & & \\
			Mean Relative Error & 2.5\% & 2.6\% & 1,344.7\%\\
			Median Relative Error  & 1.5\% & 1.7\% & 2.3\% \\
			Queries with $<$20\% Error & 99.9\% & 99.4\% & 89.4\%\\
			\bottomrule
		\end{tabular}
	}
	\caption{The relative error results when using nearest-neighbour regression with the optimizer cost as the sole feature.}
	\label{tbl:Knn}
\end{table}

Table \ref{tbl:Knn} shows the results when using nearest-neighbour regression with just the optimizer cost. We considered different values of $k$ in the range 3 to 9 but found little difference between them. We report the results for $k=5$ neighbours (the default value in scikit-learn).

The results show that switching to just the optimizer cost to identify nearest neighbours does not result in any loss of accuracy. The median relative error is very low for all our benchmarks and almost all queries in the TPC-H benchmark and over 75\% in the TPC-DS benchmark were predicted to within 20\% of their true values.

To compare to published results: we found a mean relative error for uniform TPC-H of approximately 1.7\% in our experiments which is similar to the 2.1\% reported by Akdere \textit{et al.} using the method of Ganapathi \textit{et al.}  For TPC-DS, the only published results used a 1GB benchmark, and for that benchmark Ganapathi \textit{et al.} reported that they could predict 85\% of queries to within 20\% of their true values. For the same size TPC-DS benchmark we found that our method produced the same result (84\% to within 20\%).

The TPC-DS benchmark shows significantly more variation at 10GB than at 1GB (see Section \ref{sec:Benchmarks}) and this explains why the proportion of queries successfully predicted falls. Repeating the method of Ganapathi \textit{et al.} with the 10GB TPC-DS benchmark, we found that 77\% of queries could be predicted to within 20\% of their true value which is the same as with our method.

At the operator level, results are improved further. The mean and median relative errors increase a little but the proportion of queries with less than 20\% error is increased. For the two TPC-H benchmarks over 99\% of queries have small errors and almost 90\% of TPC-DS queries are similarly predicted to within 20\% of their true values.

These results confirm that switching to the optimizer cost to identify nearest neighbours does not affect the accuracy of the results. Nearest-neighbour regression remains the most accurate method, at least when the query types being predicted are the same as those seen before. By avoiding dimensionality reduction we can produce predictions much faster. On our experimental machine, predictions took approximately 2 milliseconds to generate.

\subsection{Outliers}
\label{sec:Outliers}

\begin{figure}[t]
	\centering
	\includegraphics[width=\columnwidth,trim=4 4 4 4,clip]{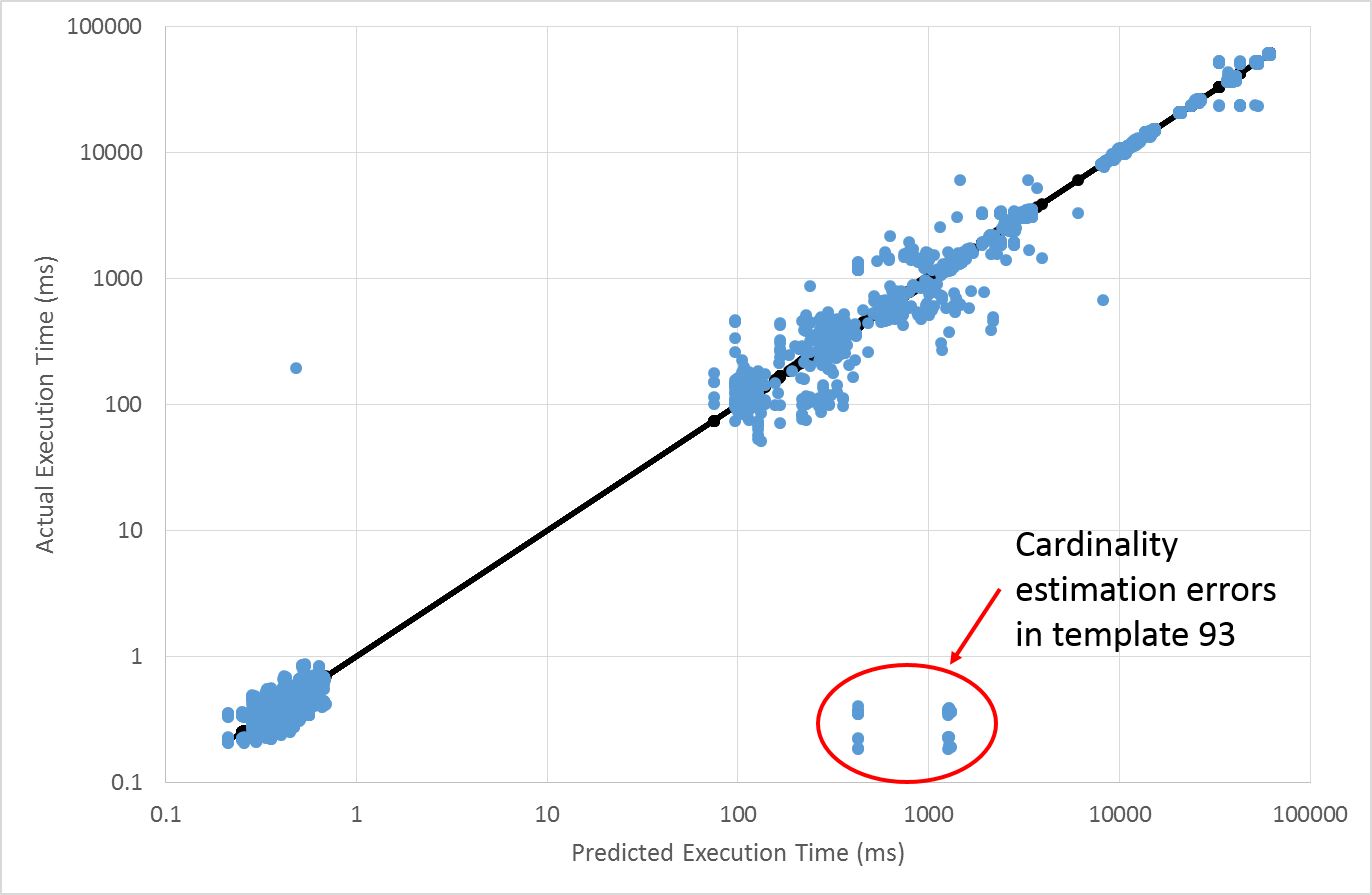}
	\caption{Plotting the real execution times against the predicted times using nearest neighbour regression for TPC-DS reveals the existence of significant outliers.}
	\label{fig:PlanLevelTPCDS}
\end{figure}

The results presented so far show the effectiveness of nearest-neighbour regression using the optimizer cost but hide the existence of a few significant outliers. Fig. \ref{fig:PlanLevelTPCDS} shows the predicted and actual execution times for the TPC-DS benchmark queries. The figure highlights a set of queries where the prediction is more than an order of magnitude away from the actual execution time. 

These queries are derived from template 93 which is a significant outlier from the benchmark. Upon closer inspection the problem is that the planner incorrectly estimates the number of rows that will be returned by a scan on one of the tables. Fig. \ref{fig:ExamplePlan} shows the query plan for an instance of template 93 and this plan is common to all instances. At the bottom of the plan is a Bitmap Scan which is the source of the problem. The planner always predicts that this scan will return approximately 61,000 rows regardless of the scan condition. In some cases this is an accurate prediction but in others the scan actually returns no rows at all. 

Since the planner incorrectly believes all instances have very similar results it assigns very similar costs to the chosen plans for every instance. In reality, however, different instances can have very different execution times. When predicting the execution time of an instance, the nearest-neighbour regression method cannot differentiate between those instances with 61,000 returned rows and those with none because they all have very similar costs. A similar problem accounts for other outliers seen in the results. 

Ultimately, the cause of the errors is the planner's inability to correctly produce significantly different query plans for queries that turn out to be significantly different. Therefore, none of the alternative QPP methods would be able to avoid making significant errors. For example, the features used by Ganapathi \textit{et al.} are a flattened version of the query plan and since the plans of different instances are virtually identical the features for the different instances will be virtually identical. Indeed Ganapathi \textit{et al.} also reported outliers resulting from poor cardinality estimates \cite{Ganapathi2009}, as did Akedere \textit{et al.} \cite{Akdere2012}.

We stated earlier that the optimizer cost's usefulness at identifying nearest neighbours is, to a large degree, independent of the accuracy of the optimizer's cardinality estimation. This is the exception - when the cardinality error is such that two predicates with different selectivities are predicted to have the same selectivity. In general, though, so long as operators with different cardinalities are given different estimates by the optimizer, nearest-neighbour regression will work regardless of whether those estimates are accurate or not.

\section{Conclusions and Future Work}

Being able to predict the execution time of queries has a number of important applications, including providing Service Level Agreements, scheduling and error detection. Two general approaches have been proposed in the literature - post-processing using machine learning and a white-box approach based on pre-processing the optimizer cost constants combined with increased sampling. Both approaches were motivated by the consensus that post-processing the optimizer cost alone is not sufficient.

We noted, however, that the evidence against post-processing the optimizer cost is limited to linear regression. In contrast, the post-processing methods in the literature which use a richer set of features rely on more complex machine learning algorithms and sometimes combine them with decomposing the query plan into its constituent operators. We have therefore revisited the issue of post-processing the optimizer cost.

We have shown that using more complex and non-linear models can improve the accuracy of predictions when using only the optimizer cost. In some cases, the accuracy is comparable to published results. For example, applying SVR at the operator level using only the optimizer cost has similar mean relative error (28.4\%) to that found when using Regression Trees at the operator level with more than 9 features per operator (26.0\%). 

Our main result, however, was to show that the most accurate method in the literature - nearest-neighbour regression - is as accurate when using only the optimizer cost as when using a larger set of features. We therefore suggest that it is too broad a statement to claim that post-processing the optimizer cost is never effective. 

Using only the optimizer cost provides the advantage of lower overheads, both in terms of storing training data and speed of prediction. On our experimental machine, predictions using nearest-neighbour regression with only the optimizer cost were made in approximately 2 milliseconds. This raises the possibility of using online learning to help mitigate the limitation of post-processing, namely that it produces poor predictions for queries that are not similar to those previously run. Although this problem is fundamental and cannot be completely avoided, if results are immediately incorporated into the training data then the poor predictions will only occur for the first one or two dissimilar queries. Thereafter, there will be similar queries in the training data from which to make accurate predictions.

We have also seen that we can accurately predict the execution time of individual operators as part of predicting the execution time of queries. Moreover, those predictions are more accurate than the ones using the analytical model. Therefore, it may be useful to replace or augment the analytical cost model with machine learning-based predictions. This would not only automatically make the cost of a plan also its predicted execution time but result in the selection of faster execution plans. 


\bibliographystyle{unsrt}
\bibliography{nearestNeighbourCost}

\begin{thebibliography}{10}

\bibitem{Gupta2008}
Chetan Gupta, Abhay Mehta, and Umeshwar Dayal.
\newblock {PQR}: Predicting query execution times for autonomous workload
  management.
\newblock In {\em Autonomic Computing, 2008. ICAC'08. International Conference
  on}, pages 13--22. IEEE, 2008.

\bibitem{Ganapathi2009}
Archana Ganapathi, Harumi Kuno, Umeshwar Dayal, Janet~L Wiener, Armando Fox,
  Michael~I Jordan, and David Patterson.
\newblock Predicting multiple metrics for queries: Better decisions enabled by
  machine learning.
\newblock In {\em Data Engineering, 2009. ICDE'09. IEEE 25th International
  Conference on}, pages 592--603. IEEE, 2009.

\bibitem{Akdere2012}
Mert Akdere, Ugur Cetintemel, Matteo Riondato, Eli Upfal, and Stanley~B Zdonik.
\newblock Learning-based query performance modeling and prediction.
\newblock In {\em Data Engineering (ICDE), 2012 IEEE 28th International
  Conference on}, pages 390--401. IEEE, 2012.

\bibitem{Li2012}
Jiexing Li, Arnd~Christian K{\"o}nig, Vivek Narasayya, and Surajit Chaudhuri.
\newblock Robust estimation of resource consumption for {SQL} queries using
  statistical techniques.
\newblock {\em Proceedings of the VLDB Endowment}, 5(11):1555--1566, 2012.

\bibitem{Wu2013}
Wentao Wu, Yun Chi, Shenghuo Zhu, Junichi Tatemura, Hakan Hacigumus, and
  Jeffrey~F Naughton.
\newblock Predicting query execution time: Are optimizer cost models really
  unusable?
\newblock In {\em Data Engineering (ICDE), 2013 IEEE 29th International
  Conference on}, pages 1081--1092. IEEE, 2013.

\bibitem{Singhal2016}
Rekha Singhal and Manoj Nambiar.
\newblock Predicting sql query execution time for large data volume.
\newblock In {\em Proceedings of the 20th International Database Engineering \&
  Applications Symposium}, pages 378--385. ACM, 2016.

\bibitem{Selinger1979}
P~Griffiths Selinger, Morton~M Astrahan, Donald~D Chamberlin, Raymond~A Lorie,
  and Thomas~G Price.
\newblock Access path selection in a relational database management system.
\newblock In {\em Proceedings of the 1979 ACM SIGMOD international conference
  on Management of data}, pages 23--34. ACM, 1979.

\bibitem{Poess2007}
Meikel Poess, Raghunath~Othayoth Nambiar, and David Walrath.
\newblock Why you should run {TPC-DS}: a workload analysis.
\newblock In {\em Proceedings of the 33rd international conference on Very
  large data bases}, pages 1138--1149. VLDB Endowment, 2007.

\bibitem{Altman1992}
Naomi~S Altman.
\newblock An introduction to kernel and nearest-neighbor nonparametric
  regression.
\newblock {\em The American Statistician}, 46(3):175--185, 1992.

\bibitem{Smola2004}
Alex~J Smola and Bernhard Sch{\"o}lkopf.
\newblock A tutorial on support vector regression.
\newblock {\em Statistics and computing}, 14(3):199--222, 2004.

\bibitem{Pedregosa2011}
F.~Pedregosa, G.~Varoquaux, A.~Gramfort, V.~Michel, B.~Thirion, O.~Grisel,
  M.~Blondel, P.~Prettenhofer, R.~Weiss, V.~Dubourg, J.~Vanderplas, A.~Passos,
  D.~Cournapeau, M.~Brucher, M.~Perrot, and E.~Duchesnay.
\newblock Scikit-learn: Machine learning in {P}ython.
\newblock {\em Journal of Machine Learning Research}, 12:2825--2830, 2011.

\bibitem{Refaeilzadeh2009}
Payam Refaeilzadeh, Lei Tang, and Huan Liu.
\newblock Cross-validation.
\newblock In {\em Encyclopedia of database systems}, pages 532--538. Springer,
  2009.

\bibitem{Council2008}
Transaction Processing~Performance Council.
\newblock {TPC-H} benchmark specification.
\newblock {\em Published at http://www. tcp. org/hspec. html}, 2008.

\bibitem{Chaudhuri}
S.~Chaudhuri and V.~R. Narasayya.
\newblock {TPC-D} data generation with skew.
\newblock Available via anonymous ftp from
  ftp.research.microsoft.com/users/viveknar/tpcdskew.

\bibitem{Moore2013}
David~S Moore, William Notz, and Michael~A Fligner.
\newblock {\em The basic practice of statistics}.
\newblock WH Freeman and Company, 2013.

\bibitem{Vert2004}
Jean-Philippe Vert, Koji Tsuda, and Bernhard Sch{\"o}lkopf.
\newblock A primer on kernel methods.
\newblock {\em Kernel Methods in Computational Biology}, pages 35--70, 2004.

\bibitem{Chang2011}
Chih-Chung Chang and Chih-Jen Lin.
\newblock {LIBSVM}: A library for support vector machines.
\newblock {\em ACM Transactions on Intelligent Systems and Technology},
  2:27:1--27:27, 2011.
\newblock Software available at \url{http://www.csie.ntu.edu.tw/~cjlin/libsvm}.

\bibitem{Hsu2003}
Chih-Wei Hsu, Chih-Chung Chang, Chih-Jen Lin, et~al.
\newblock A practical guide to support vector classification, 2003.

\bibitem{Lohman2014}
Guy Lohman.
\newblock Is query optimization a "solved" problem?
\newblock ACM Sigmod Blog, April 2014.

\end{thebibliography}

\end{document}